\documentclass{osa-article}

\journal{osajournal}
\usepackage[export]{adjustbox}
\usepackage{subfig}
\usepackage[utf8]{inputenc}
\usepackage[T1]{fontenc}

\articletype{Research Article}

\begin{document}

\title{Frequency-modulated comb in a VECSEL}

\author{Christian Kriso,\authormark{1,*} Avijit Barua,\authormark{1} Obaid Mohiuddin,\authormark{1} Christoph Möller,\authormark{1} Antje Ruiz-Perez,\authormark{1}Wolfgang Stolz,\authormark{1} Martin Koch,\authormark{1} and Arash Rahimi-Iman\authormark{1}}

\address{\authormark{1}Department of Physics and Material Sciences Center, Philipps-Universität Marburg, Renthof 5, 35032 Marburg, Germany}
\email{\authormark{*}christian.kriso@physik.uni-marburg.de} 



\begin{abstract*}
Optical frequency combs based on mode-locked lasers have revolutionized many areas of science and technology, such as precision metrology, optical frequency synthesis or telecommunications. In recent years, a particular kind of frequency comb has been observed in edge-emitting semiconductor lasers where the phase difference between longitudinal laser modes is fixed but not zero. This results in a linearly chirped output in the time domain with nearly constant intensity. Here, by using coherent beatnote spectroscopy, we show that such a comb regime can also exist in vertical-external-cavity surface-emitting lasers (VECSELs), as evidenced for a specific part of the laser spectrum. Our findings may not only lead to a better understanding of the physics of frequency-modulated combs but also enable comb applications with high optical power per comb line and flexible emission wavelengths.
\end{abstract*}

\section{Introduction}
An optical frequency comb consists of equidistantly spaced laser lines. What distinguishes it from a free-running multi-mode laser is the fact that there exists a fixed phase relationship between adjacent longitudinal modes oscillating in the laser resonator. This guarantees that the frequency spacing between different modes does not change with time. Historically, such mode-locking has been achieved by the use of strongly intensity dependent elements in the laser cavity where saturable absorption provides a coupling mechanism between laser modes, which typically results in ultrashort pulses in the time-domain and can be understood as in-phase synchronization of the laser modes \cite{Haus2000, Hillbrand2019a}. However, frequency-comb generation in a semiconductor laser can also occur in the absence of a saturable absorber \cite{Hugi2012a, Escoto2020}. In this case, it is understood that the beating of different laser modes leads to oscillations of the carrier density in the gain medium at multiples of the repetition frequency (corresponding to the longitudinal mode spacing) which in turn again couple with the laser modes. This provides a locking mechanism that can compensate dispersion and noise that otherwise prevents equidistance and phase coherence of the modes, respectively. These combs have been primarily observed in interband and quantum cascade lasers, which have a short gain recovery time compared to the cavity round trip time, as a slow gain medium with a long gain recovery time cannot respond efficiently to the fast intermode beating and thus results in negligible mode-coupling \cite{Piccardo2019}. However, also in interband diode lasers, where the round trip time is typically somewhat smaller than the carrier lifetime of around 1 ns, this effect has been observed in a variety of material systems \cite{Day2020, Rosales2012,Hillbrand2019a}. This has been supported by recent theoretical work which suggests that comb formation is principally determined by the interplay of group delay dispersion (GDD) and carrier-induced refractive index changes leading to soliton-like states with a characteristic linear chirp or frequency-modulation (FM) in the time-domain \cite{Opacak2019, Burghoff2020}. \\
Optically pumped semiconductor disk lasers or VECSELs can be passively mode-locked by use of a semiconductor-saturable-absorber-mirror (SESAM) with ultrashort pulse emission down to the sub-100-fs regime \cite{Waldburger2016, Laurain2018}, and even saturable-absorber free mode-locking was discussed \cite{Kornaszewski2012, Albrecht2013,Gaafar2014, Gaafar2016}. Whether this class of lasers support FM combs similarly to their edge-emitting counterparts remains an exciting question to address, as both types of semiconductor lasers differ in some aspects fundamentally. The mode spacing (i.e. the free spectral range), which directly relates to the cavity round trip time, is typically an order of magnitude smaller, in the few GHz range. This provides a setting, where gain recovery time and round trip time are very similar. Also, the spatial hole burning process of a Fabry-Pérot cavity, that triggers multi-mode operation in edge-emitting lasers, is absent in VECSELs as the gain-contributing quantum wells only overlap at nearly discrete positions in the sample structure with the standing optical field. In addition to that, the typically centimeter-long external resonator leads to the storage of optical power away from the gain medium. 

From a practical point of view, FM combs in VECSELs might be useful for dual-comb spectroscopy with high power per comb line and consequently increased SNR of the measurement, as they are unrivaled in terms of output power among semiconductor lasers, reaching more than 100 W in cw operation \cite{Heinen2012}. Moreoever, advanced III-V semiconductor technology enables great flexibility with respect to the emission wavelength, ranging from the visible to the mid-infrared \cite{Rahimi-Iman2016}. Recently, dual-comb spectroscopy with mode-locked VECSELs has been demonstrated \cite{Link2017}. However, when pushing the pulse length below 100 fs one observes a dramatic decrease in peak and average power \cite{Waldburger2016}, which is possibly caused by fundamental non-equilibrium carrier dynamics like kinetic hole burning \cite{Kilen2014}. For increased average power and thus more power per comb line one would have to increase the pulse length, which means generating chirped pulses for a given bandwidth. A frequency-modulated comb, being the extreme case of a maximally linearly-chirped "pulse", would lead to quasi-cw operation and thus highest average power. Remarkably, the existence of FM combs in lasers with fast gain media has been explained with the so-called maximum-emission-principle, which states that the phase\textendash amplitude relations of the laser modes will organize in a way to extract the maximum amount of power from the gain medium which corresponds to the FM state \cite{Piccardo2019}. \\
In this work, we investigate whether FM combs can also be supported in VECSELs by using a coherent beatnote spectroscopy technique, referred to as SWIFTS (Shifted Wave Interference Transform Spectroscopy)\cite{Burghoff2015}. This allows us to directly measure the intermode phase relation of the laser and assess its phase stability (coherence) over the laser spectrum. Our results show that such comb states can indeed exist in VECSELs and we discuss its spectral and pump-power dependence.

\section{Sample properties and experimental setup}
\begin{figure}[h!]
 \centering
    \subfloat{\includegraphics[width=6cm,valign=c]{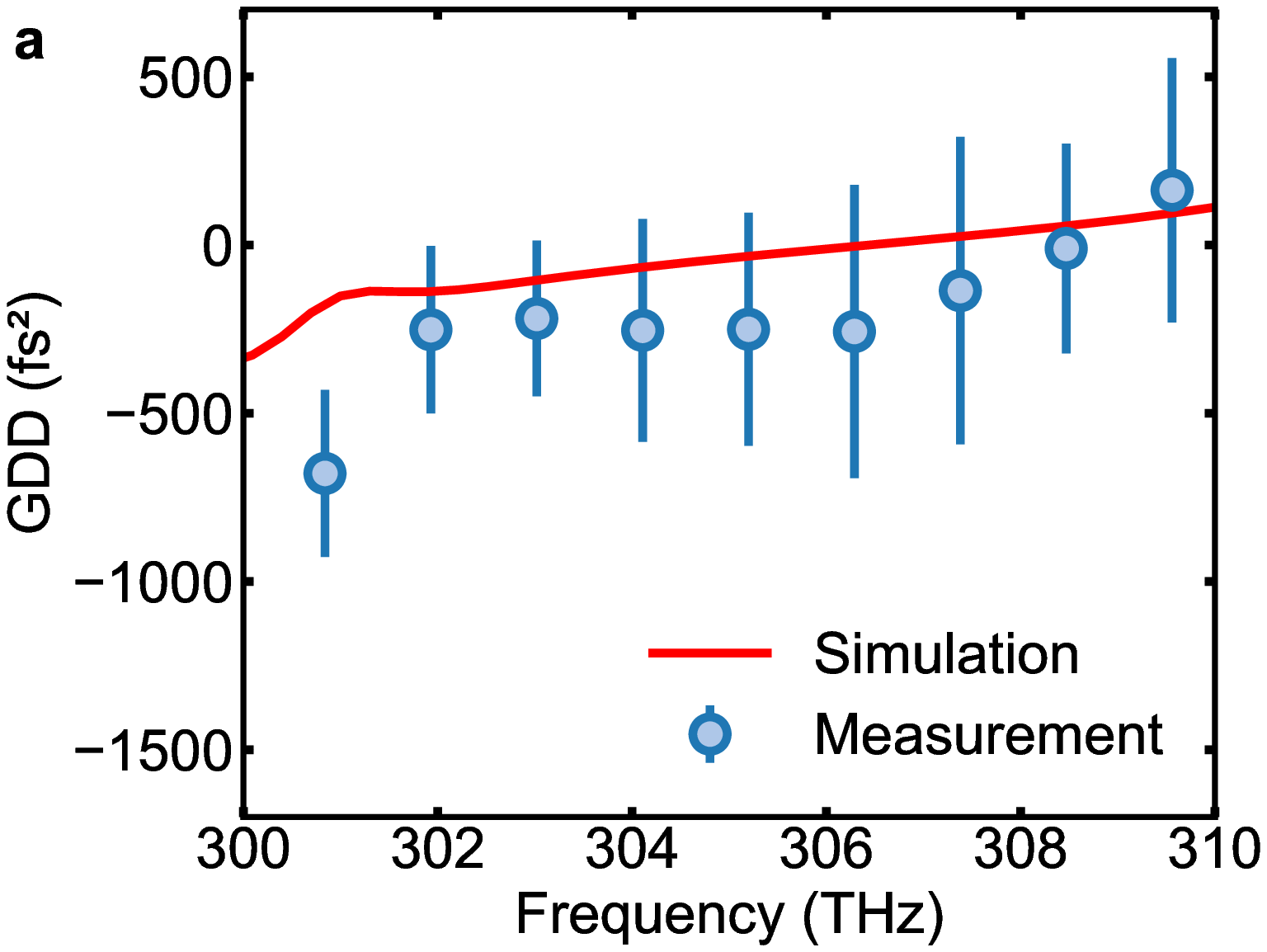}}%
    \qquad
    \subfloat{{\includegraphics[width=6cm,valign=c]{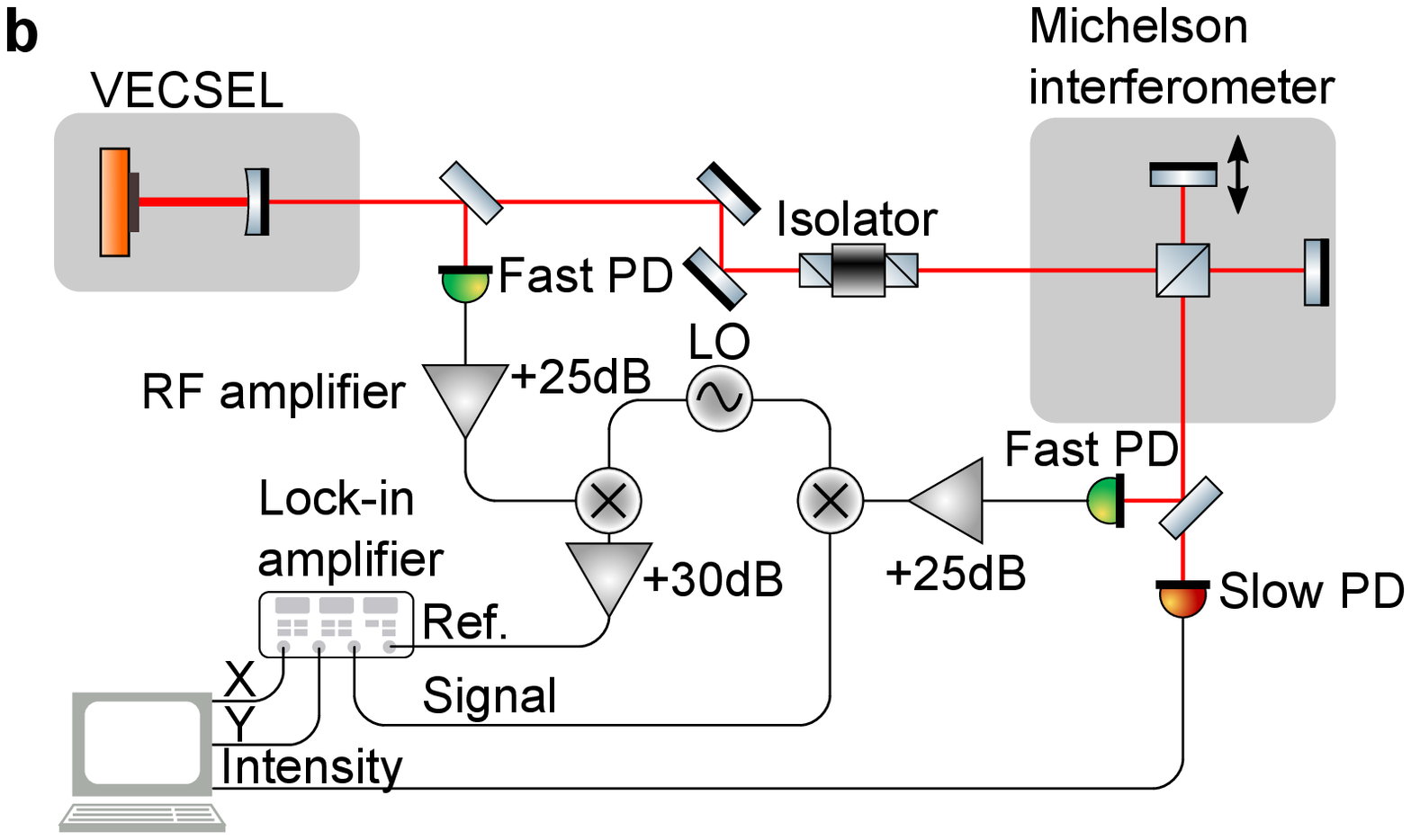}}}%
    \caption{(a) Measured and simulated group delay dispersion (GDD) of the investigated sample. The errorbars indicate the standard deviation when averaging over 997 subsequent measurements/interferograms. (b) Experimental setup used for coherent beatnote spectroscopy. PD stands for photodiode, LO for local oscillator.}%
    \label{Dispersion_and_setup}%
\end{figure}

For our investigations, we use a VECSEL gain chip, which has been designed for ultra-short pulse emission and low dispersion. To ensure a good crystalline quality, the VECSEL layer structure was grown by metal organic vapor phase epitaxy (MOVPE) using an Aixtron AIX 2600 G3 reactor.
The active laser structure consists of four double (GaIn)As quantum wells (QWs) with surrounding Ga(AsP) strain compensating barriers. Each QW-pair was placed around a maximum of the longitudinal optical field. The distributed Bragg reflector contains 23.5 (AlGa)AlGaAs mirror pairs with $\lambda$/4 layers for the VECSEL lasing wavelength at 980~nm and additional 13 pairs for the pump wavelength at 808~nm to increase the pump efficiency. The final GaAs layer is for phase matching and oxidation protection. The detailed chip structure and layer composition are provided in the supplementary material.\\
We have performed measurements with a white-light Michelson interferometer to determine the group delay dispersion (GDD) of the sample \cite{Gosteva2005}. As can be seen in Fig. \ref{Dispersion_and_setup}(a), the GDD at the lasing frequencies around 303~THz is flat and lower than 500~fs$^2$ in magnitude as expected for an anti-resonant design, i.e. a design that suppresses a strong microcavity resonance for optimized broadband operation at the cost of reduced gain \cite{Tropper2006}. Also, the measured dispersion is close to the calculated dispersion obtained by transfer-matrix simulations of the layered chip structure. More details on the dispersion measurement and calculation are provided in the supplementary material.

For laser operation, we mount the chip on a temperature-controlled copper heat sink and construct a simple linear cavity of approximately 9~cm length with an output mirror of 100 mm radius of curvature, minimized dispersion and 0.6 \% outcoupling rate. The chip is pumped with a fiber-coupled 808-nm multi-mode diode laser and the pump-spot diameter is adjusted to 200~µm to match the fundamental transverse mode on the chip. The temperature of the heat sink is kept at 18°C throughout all investigations.

In order to investigate whether the modes of the laser are phase-stable and locked, we use the SWIFTS technique, which resolves the beatnote coherently over the laser spectrum by use of a Michelson interferometer \cite{Burghoff2015}. A sketch of our experimental setup is shown in Fig. \ref{Dispersion_and_setup}(b). Part of the laser light is branched off into a fast photodiode while the other part is sent into a home-built Michelson interferometer, the output of which is recorded by another fast photodiode. The signal of both photodiodes is subsequently amplified (by around 25 dB) and mixed down with a local oscillator (LO) with a frequency of 1.58 GHz, which is close to the fundamental beatnote frequency of 1.6 GHz. The downconverted signal of the first photodiode is again amplified (by around 30 dB) to serve as reference signal for a fast lock-in amplifier (SR844, Stanford Research Systems). The downconverted signal of the fast photodiode after the Michelson interferometer is taken as signal input. Consequently, the lock-in amplifier serves as phase-sensitive detector, which records two interferograms when the arm of the Michselon interferometer is scanned. One interferogram, $X(\tau)$, corresponds to the in-phase component with respect to the reference signal and the other interferogram, $Y(\tau)$, corresponds to the quadrature (90°-shifted) component. It can be shown (see the supplementary material) that, by Fourier-transforming these interferograms with respect to the interferometer delay, the intermode phase difference (SWIFTS phase) can be retrieved by $\arg(X(\omega)-iY(\omega))$, where $X(\omega)$ and $Y(\omega)$ are the complex spectra of $X(\tau)$ and $Y(\tau)$ \cite{Burghoff2015, Hillbrand2019a}.  The magnitude of the combined spectra, the SWIFTS intensity spectrum $|X(\omega)-iY(\omega)|$, indicates over which spectral regions the modes are locked. By comparing this spectrum to the normal intensity spectrum obtained by Fourier-transforming the interferogram recorded by a slow photodiode, one can assess which parts of the spectrum are phase-coherent, and to what degree. When the shape of both spectra matches exactly, the emission consists of perfectly phase-coherent and equidistant laser modes.

\section{Experimental results}
\begin{figure}[h!]
\centering\includegraphics[width=13cm]{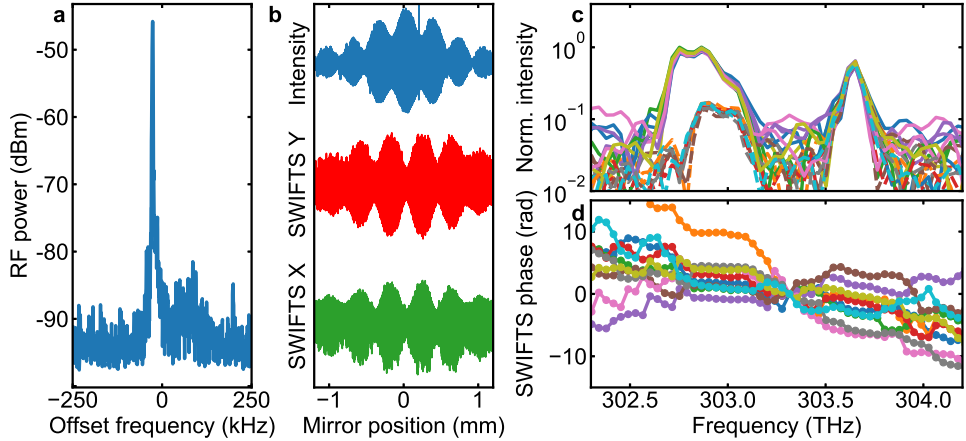}
\caption{Coherence characterization of the VECSEL at 3.8 W pump power. (a) Fundamental beatnote at 1.6 GHz recorded with a resolution bandwidth of 1 kHz over a span of 500 kHz. (b) Intensity interferogram recorded with a slow photodiode ("Intensity") and the interferograms of the in-phase ("SWIFTS X") and quadrature component ("SWIFTS Y") of the lock-in detection. (c) Magnitude of the Fourier-transforms of the intensity interferogram (solid lines) and the combined SWIFTS interferograms (dashed lines). Ten subsequently recorded measurements are shown. (d) Corresponding frequency-resolved intermode (SWIFTS) phase. The dots indicate the resolution of the measurement and do not represent individual modes.}
\label{fig:example_SWIFTS}%
\end{figure}

A first indicator of phase-locking in a laser is provided by a narrow RF beatnote.  
Figure \ref{fig:example_SWIFTS}(a) shows the fundamental RF beatnote measured at 1.6~GHz with a 25~GHz photodiode at an optical pump power of 3.8~W. The 3-dB-width of the beatnote is below the 1~kHz resolution bandwidth of the spectrum analyzer. The noise pedestal is nearly 30~dB below the peak.
Figure \ref{fig:example_SWIFTS}(b) displays the envelope of the raw interferograms recorded with the slow photodiode and the SWIFTS setup. The regular intensity interferogram exhibits a maximum at zero delay of the interferometer. The beating on the envelope is caused by the two-color operation of the laser. In constrast, both SWIFTS interferograms show a characteristic minimum at zero delay. This minimum is attributed to the presence of frequency-modulation in the laser \cite{Hugi2012a} and can be understood by regarding the intermode beating as complex phasors that cancel each other out when summed up \cite{Hillbrand2019a}.
Finally, Fig. \ref{fig:example_SWIFTS}(c) shows the retrieved normalized intensity and the two SWIFTS spectra as well as the intermode (SWIFTS) phase for 10 subsequent measurements (Fig. \ref{fig:example_SWIFTS}(d)). The fact that the laser operates on two spectral lobes is probably a consequence of the anti-resonant design of the gain chip, for which the cavity resonance is designed in a way that there is no strong cavity resonance at one specific wavelengths but rather two smaller resonances centered around the maximum material gain (see the supplementary material). Interestingly, for the lobe at larger frequencies, the intensity and SWIFTS spectrum match very well, whereas for the lobe at smaller frequencies, the amplitude of the SWIFTS spectrum is considerably smaller than the intensity spectrum and does not span over the whole range towards smaller frequencies. Accordingly, the SWIFTS spectrum is here normalized with respect to the intensity of the higher-frequency lobe. The intermode phases of the lobe at larger frequencies are linearly chirped, while the slope for the phases at the lower frequency lobe is rather flat. Note that the relative phase offset between the modes of the two lobes cannot be unambiguously determined due to the lack of spectral intensity between the two lobes. Still, the slope of the intermode phases are well reproduced in subsequent measurements, as well as the intensity and SWIFTS spectra, and therefore we legitimately average over multiple subsequent measurements in the following investigations.

\begin{figure}[h!]
 \centering
    \subfloat{\includegraphics[width=6.5cm,valign=c]{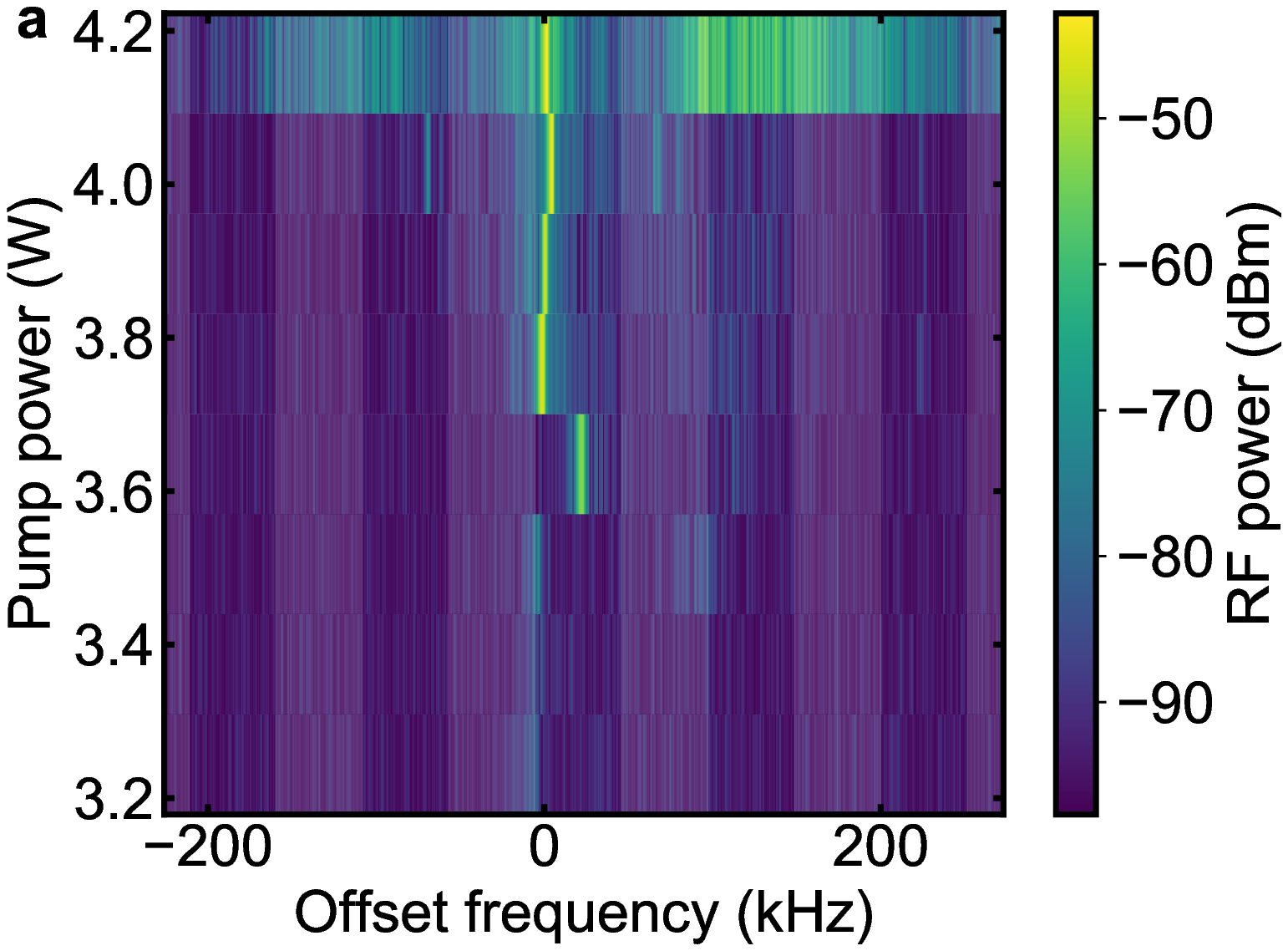}}%
    \qquad
    \subfloat{{\includegraphics[width=5.5cm,valign=c]{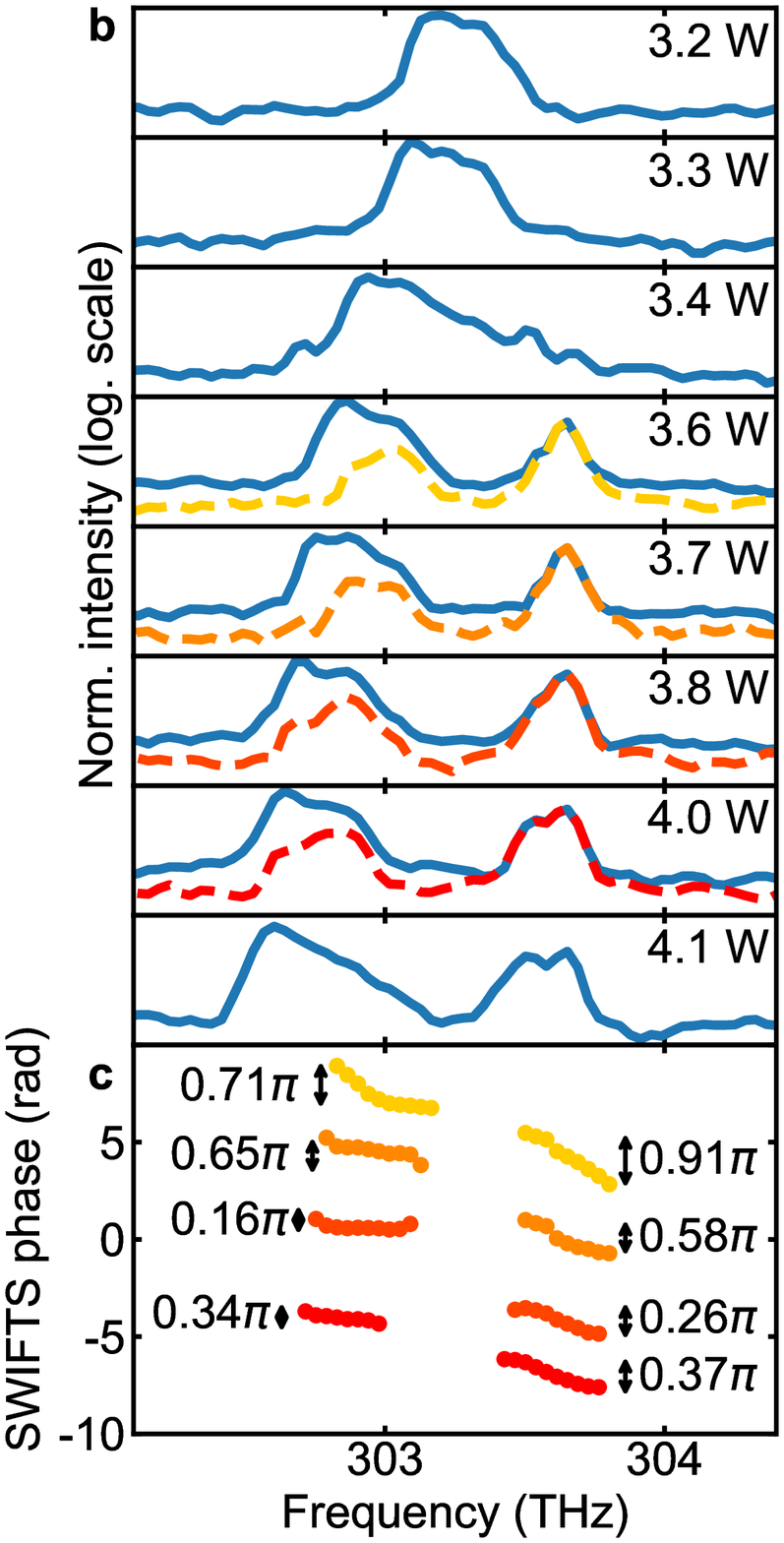}}}%
    \caption{(a) Pump-power dependence of the beatnote at 1.6~GHz measured with 1~kHz resolution bandwidth and 500~kHz span. (b) Pump-power dependent intensity spectra (blue solid line) and SWIFTS intensity spectra (dashed line). The respective pump-power value is indicated in each subfigure. (c) Corresponding SWIFTS phase. The power level is displayed color-coded as in the above SWIFTS intensity spectra. The dots indicate the spectral resolution. Pump power increases from top to bottom. The total phase differences of the phase spectra of individual lobes are indicated in the figure. }%
    \label{fig:RF_and_pd_SWIFTS}%
\end{figure}

To gain further insights into the coherence properties of our laser, we investigate the pump power dependence of the fundamental beatnote, as well as of the intensity and SWIFTS spectra. In Fig. \ref{fig:RF_and_pd_SWIFTS}(a), the pump-power-dependent evolution of the beatnote at 1.6~GHz is displayed. At 3.2~W pump power, corresponding to an output power of the laser of 111~mW, a detectable but small beatnote close to the noise level is measured. At pump powers ranging from 3.2 to 3.6~W, the beatnote is so small, that it cannot be used as reference for the SWIFTS measurements. However, at a pump power of 3.6~W, the beatnote suddenly increases by more than 20 dB in magnitude. Strikingly, this coincides with the separation of the optical spectra in Fig. \ref{fig:RF_and_pd_SWIFTS}(b) into two lobes. Here, as already observed in Fig. \ref{fig:example_SWIFTS}(b), the intensity and SWIFTS spectra of the lobe at larger frequencies match perfectly over the pump power range from 3.4 to 4~W corresponding to output powers of 135 to 188~mW. In this range, the lobe at lower frequencies only exhibits partial coherence with nearly no coherence towards the lower frequency side. At a pump power of 4.1~W, the beatnote "explodes" into strong and broad noise pedestals around its peak within the frequency range of 500~kHz. The phase coherence is lost and the SWIFTS intensity spectrum consequently drops to zero. For larger pump powers, the beatnote remains broad. Thus, no data for pump powers beyond 4.1~W are shown.\\
When looking at the retrieved intermode phases (lower part of Fig. \ref{fig:RF_and_pd_SWIFTS}(b)), one observes that the intermode phase of the higher-frequency lobe exhibits a larger linear chirp than the phase of the low-frequency lobe. In both cases, the slope tends to decrease with increasing pump power. At a pump power of 3.4~W, the total phase difference over the two lobes sums up to 1.62$\pi$ which is relatively close but not equal to the phase difference of 2$\pi$ expected for a frequency-modulated comb with a continuous spectrum \cite{Hillbrand2019a, Opacak2019, Singleton2018}. We presume that the interplay of the modes in the two lobes leads to a more complex synchronization phenomenon than anti-phase synchronization, for which the intermode phases are splayed exactly over the range of 2$\pi$. Further work is required to complement our experimental findings towards a more complete picture of the laser properties and the observed amplitude\textendash phase-locking behavior.

\begin{figure}[h!]
\centering\includegraphics[width=13cm]{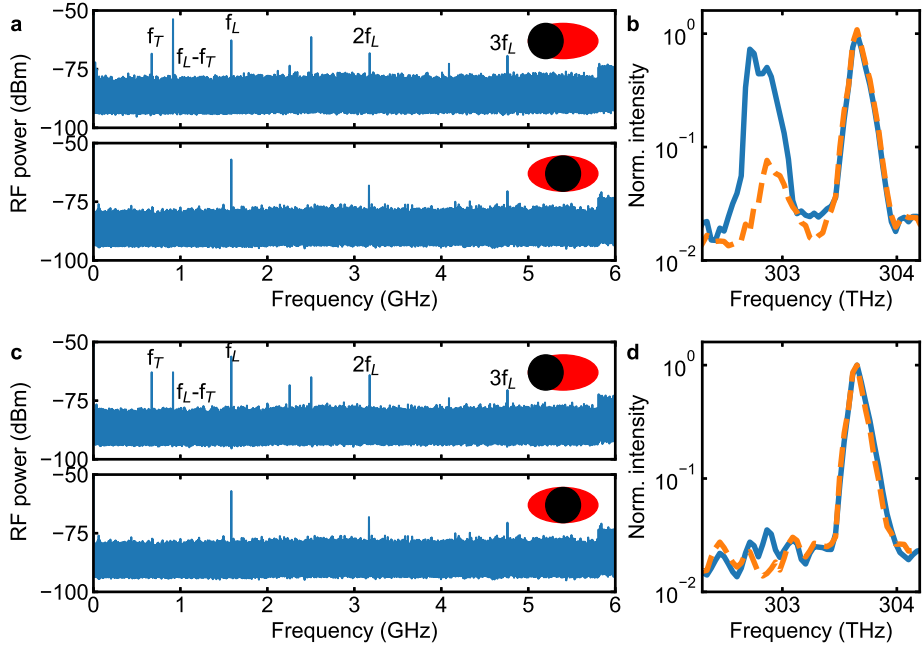}
\caption{Effect of spectral and spatial filtering on the long-span RF spectrum at a pump power of 3.3~W (different sample spot than results of Fig. \ref{fig:RF_and_pd_SWIFTS})): (a) Long-span RF spectrum (resolution bandwidth 100 kHz) of unfiltered laser emission for two positions of the photodiode with respect to the optical beam, indicated by the sketch in the upper right corner. The red ellipse visualizes the laser beam and the black circle the photodiode. (b) Intensity (blue solid line) and SWIFTS spectrum (dashed orange line) of unfiltered laser emission. (c) Long-span RF spectrum (resolution bandwidth 100~kHz) of filtered laser emission for two positions of the photodiode.(d) Intensity (blue solid line) and SWIFTS spectrum (dashed orange line) of filtered laser emission.}
\label{fig:PD_position_dependent_RF_spectra}%
\end{figure}

It is worth mentioning that one witnesses here the existence of one higher order transverse mode with a broad RF line in the long-span RF spectrum as shown in Fig. \ref{fig:PD_position_dependent_RF_spectra}(a). Attempts to suppress this mode by reducing the pump spot size on the chip were not successful. In fact, the existence of a higher order transverse mode seems to coincide with the appearance of a narrow beatnote for the fundamental mode, similar to the behavior observed in Ref. \cite{Tsou2015}. This might hint to a possible connection between the frequency-modulated comb state we observe here and the self-mode-locking reports for VECSELs, where coherence of the laser emission was evidenced by autocorrelation measurements \cite{Tsou2015, Gaafar2014}. Nonetheless, we point out that the excellent correspondence of SWIFTS and intensity spectrum for the high-frequency lobe demonstrates that this part of the spectrum must principally consist of phase-locked longitudinal modes of the fundamental mode. \\
Furthermore, we investigated the effect on the RF spectrum of suppressing the partially incoherent lobe at lower frequencies with an angle-tunable short-pass filter (Semrock TSP01-1116). Here, it is important to note that the position of the photodiode with respect to the beam cross section strongly influences whether lines corresponding to a higher-order transverse mode appear in the RF spectrum. To account for this effect, we recorded two long-span RF spectra for the unfiltered and filtered laser emission, respectively, one with placing the photodiode in a way to maximize the signal of the transverse mode in the RF spectrum and another with minimizing its effect. In Fig. \ref{fig:PD_position_dependent_RF_spectra}a one sees these two RF spectra for the unfiltered laser output. The upper spectrum displays a line at $f_T$=663 MHz, which corresponds to the first higher-order transverse mode (see Ref. \cite{Tsou2015}). While the signal at $f_T$ is smaller than the signal at the fundamental repetition frequency $f_L$ at 1.6~GHz, the beating product of the two modes at $f_L$-$f_T$=937~MHz dominates the RF spectrum. When placing the photodiode in a way to minimize the effect of the higher-order mode, its contribution to the spectrum disappears (lower part of Fig. \ref{fig:PD_position_dependent_RF_spectra}(a)). \\
If the low-frequency mode is suppressed with the short-pass filter, the RF line of the fundamental mode at $f_L$ (and its higher harmonics) dominate the RF spectrum. This demonstrates that mainly the low-frequency lobe is contributing to the higher-order mode. The fact that, even with filtering, the effect of the higher-order mode can be seen in the RF spectrum might be attributed to the fact that there is still a residual, incoherent spectral component below 303 THz in the filtered optical spectrum, as can be seen in Fig. \ref{fig:PD_position_dependent_RF_spectra}(d). A shortpass filter with a stronger slope and suppression ratio might eliminate the effect of the higher-order  mode completely at any position of the photodiode. The lobe in  Fig. \ref{fig:PD_position_dependent_RF_spectra}(d) has a 3-dB-width of around 0.14~THz, thus containing more than 88 longitudinal laser modes, which is more than double the number of modes that have been used in the pioneering first dual-comb spectroscopy experiment with VECSELs \cite{Link2017}. This highlights the importance of our results. Even though not the whole unfiltered laser spectrum is perfectly phase-locked, a combination of spatial and spectral filtering can render it effectively coherent.\\
We believe that with a detailed understanding of the locking mechanism it will be possible in the future to obtain FM combs in VECSELs that are fully coherent over the whole laser spectrum. Then they might provide a very simple and at the same time high-power source for dual-comb spectroscopy. To reach this goal, two open questions need to be addressed in future works: What limits the coherence over a broad spectrum and what is the optimal amount of GDD in the cavity (see Ref.\cite{Opacak2019})?\\
Moreover, FM combs may become a useful alternative to SESAM-mode-locked VECSELs for some applications, particularly in the context of novel material systems targeting new spectral emission wavelengths where SESAM mode-locking might be difficult to achieve. An example can be seen in the recently demonstrated type-II VECSELs \cite{Moeller2016}, which have not been SESAM-mode-locked yet. 
\section{Conclusion}
By measuring the intermode phase relation and the phase coherence with the SWIFTS technique, we have demonstrated that a VECSEL can run in a frequency-modulated regime, whereas at the moment this occurs over a certain part of its spectrum. This establishes optically pumped semiconductor disk lasers as a new platform where the physics of FM combs can be studied. In fact, VECSELs offer unique characteristics such as a very similar gain recovery time and cavity round trip time as well as the absence of spatial hole burning as driver for multi-mode emission. Given future optimization efforts, it can be anticipated that these comb sources may prove themselves useful for dual-comb spectroscopy in application scenarios where a particular high power per comb line or specific target wavelengths are required.

\section*{Funding}
Funding by the German Research Foundation (Deutsche Forschungsgemeinschaft) under Grant DFG RA 2841/1-1 and RA 2841/1-3 is gratefully acknowledged.

\section*{Acknowledgments}
We thank C. Schindler for expert design of the RF amplifiers and mixers and for useful discussions.

\section*{Disclosures} The authors declare no conflict of interests.
\section*{Supplementary material}

\section*{S1: Sample structure}
\begin{figure}[h!]
\centering
\fbox{\includegraphics[width=10cm]{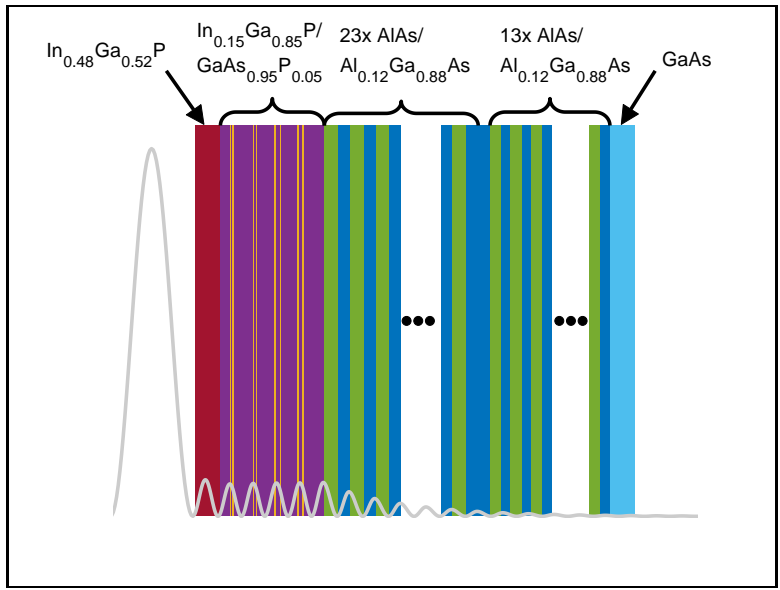}}
\caption{Sample structure and normalized standing electric field intensity at 303.65 THz (987.3 nm).}
\label{fig:sample_structure}
\end{figure}

\section*{S2: Dispersion measurement}
\begin{figure}[h!]
\centering
\fbox{\includegraphics[width=13cm]{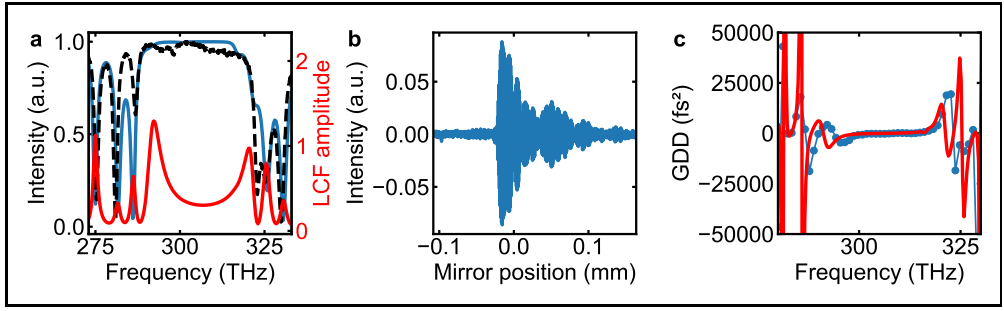}}
\caption{(a) Calculated (blue solid line) and measured (black dashed line) reflectivity spectrum of the studied VECSEL chip. Additionally, the calculated longitudinal confinement factor (LCF) of the chip's microcavity is shown (red solid line). (b) Exemplary white-light interferogram of the probed VECSEL chip. (c) Measured (blue dots) and calculated (red line) group delay dispersion (GDD). }
\label{fig:dispersion_supplement}
\end{figure}
For the measurement of the group delay dispersion (GDD) of the unpumped VECSEL chip, a home-built white-light interferometer was used, similar to the ones in Refs. \cite{Diddams1996}  and \cite{Gosteva2005}. A He-Ne laser was used as position reference of the moving mirror, mounted on a mechanical shaker, and spatially overlapped with the white-light from an halogen lamp. After the interferometer, the He-Ne laser and the white light are separated by appropriate dichroic mirrors and recorded each with a photodiode. A fast digitizer card records around 1000 interferograms while the shaker is oscillating. An examplary interferogram is shown in Fig. \ref{fig:dispersion_supplement}(b). The  mirror position has been obtained by taking the zero-crossings of the He-Ne laser interferogram as position reference. The GDD is then obtained by Fourier-transforming it with respect to the delay $\tau = \Delta x/c$, with $\Delta x$ the mirror position and $c$ the speed of light. The phase of the obtained spectrum is then twice numerically derived with respect to the angular frequency $\omega$ which provides the GDD. For averaging, multiple interferograms are interpolated over the same mirror position/delay grid before Fourier-transforming each of them and subsequently averaging the calculated GDD. Finally, the GDD of the balanced interferometer with two metal mirrors (and the sample removed) is recorded to account for the interferometer's transfer function. Figure \ref{fig:dispersion_supplement}(c) shows the measured dispersion of the sample obtained by averaging over 997 interferograms (blue dots) and shows good correspondence to the calculated dispersion. The calculated dispersion has been obtained by calculating the complex reflectivity spectrum of the sample structure with the transfer matrix formalism detailed in \cite{Tropper2006} and obtaining the second derivative of the electric field's phase with respect to $\omega$. The calculated and measured reflectivity spectra show good correspondence (see Fig. \ref{fig:dispersion_supplement}(a)). Also shown in Fig. \ref{fig:dispersion_supplement}(a) is the calculated longitudinal confinement factor (LCF) that characterizes the overlap of the intensity of the electric field with the quantum wells (see Ref. \cite{Tropper2006}).  It is at its minimum in the spectral region where the laser operates, which yields the desired flat and low dispersion, but reduces the achievable maximum output power of the sample.

\section*{S3: SWIFTS measurement}
To understand the working principle of the SWIFTS measurement (see also Refs. \cite{Burghoff2015, Burghoff2018}), we write the electric field of the laser emission as
\begin{equation}
E(t) = \sum_{n}E_n e^{i(\omega_n t+\varphi_n)} + c.c.,
\label{eq:E_field_laser}
\end{equation}
where $E_n$ is the amplitude, $\varphi_n$ the phase and $\omega_n$ the oscillation frequency of an individual mode $n$. When detecting this field with a fast photodetector after a Michelson interferometer, which introduces a delay $\tau$ in one of its arms, the detected signal becomes
\begin{equation}
\begin{split}
I(t,\tau)=\frac{1}{2}(E(t)+E(t-\tau))^2=\frac{1}{2}\sum_{n,m}(E_n E_m e^{i((\omega_n-\omega_m)t+\varphi_n-\varphi_m)}\\ +E_nE_m e^{i((\omega_n-\omega_m)(t-\tau)+\varphi_n-\varphi_m)}+2E_nE_me^{i((\omega_n-\omega_m)t+\omega_m\tau+\varphi_n-\varphi_m)})+c.c.,
\end{split}
\label{eq:PD_term_after_MI}
\end{equation}
where we have omitted any terms with an $e^{i(\omega_n+\omega_m)t}$-dependency, which will not be detected by a photodiode. Equation \ref{eq:PD_term_after_MI} can be rewritten as 
\begin{equation}
I(t,\tau)=\frac{1}{2}\sum_{n,m}E_n E_me^{i((\omega_n-\omega_m)t+\varphi_n-\varphi_m)}(1 \\ +e^{i(\omega_n-\omega_m)\tau}+2e^{i\omega_m\tau})+c.c. .
\label{eq:PD_term_after_MI_rewritten}
\end{equation}
A slow photodiode will only detect terms with $n=m$, which leads to 
\begin{equation}
DC(\tau)=\sum_{n}E_n^2(1+e^{i\omega_n\tau})+c.c. .
\label{eq:DC_interferogram}
\end{equation}
This interferogram provides the optical intensity spectrum of the laser when Fourier-transformed. \\ \\
In SWIFTS, the detected signal in Eq. \ref{eq:PD_term_after_MI_rewritten} is mixed with an in-phase ($\cos(\omega_0t)$) and a quadrature signal ($\sin(\omega_0t)$) at a frequency $\omega_0$.
This operation, which is performed by the lock-in amplifier, leads to two signals,
\begin{subequations}
\begin{eqnarray}
 X(t,\tau)= \frac{1}{4}\sum_{n,m}E_n E_m(e^{i((\omega_n-\omega_m-\omega_0)t+\varphi_n-\varphi_m)}+e^{i((\omega_n-\omega_m+\omega_0)t+\varphi_n-\varphi_m)})(...)+c.c., \\
  Y(t,\tau)= \frac{i}{4}\sum_{n,m}E_n E_m(e^{i((\omega_n-\omega_m-\omega_0)t+\varphi_n-\varphi_m)}-e^{i((\omega_n-\omega_m+\omega_0)t+\varphi_n-\varphi_m)})(...)-c.c. . 
\end{eqnarray}
\label{eq:SWIFTS_interferograms_time-dependent}
\end{subequations}
Only variations in the order of the time constant of the lock-in amplifier will be detected, which means that only $\omega_0 = \omega_n-\omega_m$ will result in a non-zero signal. 
For the measurement of the phase and coherence of two adjacent modes,  $\omega_0 = \omega_{m+1}-\omega_m$ is chosen. Note, however, that the time constant does not set the condition for the equidistance of two laser lines but just blocks lines other than nearest-neighbor in the lock-in detection. The precision of the equidistance assessment is instead determined by the inverse of the total measurement time of the interferogram (around 1 minute in our case) as pointed out in Ref. \cite{Burghoff2018}. Remarkably, this means that the equidistance measurement exhibits sub-Hz precision.
In the self-referenced scheme, as done in this work, the frequency $\omega_0$ of the reference signals is directly obtained from the fundamental beatnote of the laser, measured before the Michelson interferometer with another fast photodiode. Note that, as in our case the bandwidth of the lock-in amplifier is lower (200 MHz) than the fundamental repetition rate of the laser (1.6 GHz), we use a local oscillator and RF mixers to down-convert the signals from the fast photodiodes. \\
After the lock-in detection, Eqs. \ref{eq:SWIFTS_interferograms_time-dependent} thus become
\begin{subequations}
\begin{eqnarray}
 X(\tau)= \frac{1}{4}\sum_{m}E_{m+1} E_me^{i(\varphi_{m+1}-\varphi_m)}(...)+c.c. . \\
  Y(\tau)= \frac{i}{4}\sum_{m}E_{m+1} E_me^{i(\varphi_{m+1}-\varphi_m)}(...)-c.c. . 
\end{eqnarray}
\label{eq:SWIFTS_interferograms}
\end{subequations}
The two SWIFTS interferograms (in-phase and quadrature) can be combined in one complex interferogram,
\begin{subequations}
\begin{eqnarray}
 X(\tau)-iY(\tau)= \frac{1}{2}\sum_{m}E_{m+1} E_me^{i(\varphi_{m+1}-\varphi_m)}(1 +e^{i\omega_0\tau}+2e^{i\omega_m\tau}).
\end{eqnarray}
\label{eq:combined_SWIFTS_interferograms}
\end{subequations}
One can see that the product of two adjacent mode amplitudes and their phase difference will be resolved over the optical spectrum when the interferograms are Fourier-transformed with respect to the delay $\tau$. Now it is obvious that $\arg(X(\omega)-iY(\omega))$ will provide the intermode phase $\varphi_{m+1}-\varphi_{m}$. However, the contribution of two adjacent modes to the SWIFTS spectrum $|X(\omega)-iY(\omega)|$ will disappear when their phases fluctuate over time scales of the total measurement time. Thus, the comparison of $|X(\omega)-iY(\omega)|$ with the DC intensity spectrum provides a measure of coherence (i.e. phase stability) of the laser.


\bibliography{bib_SWIFTS_VECSEL}

\end{document}